\documentclass[article,showpacs,amsmath,amssymb]{revtex4}
\usepackage{graphicx}
\usepackage{dcolumn}
\usepackage{bm}

\begin{document}

\title{Random walk on ${\bf p-}$adics in glassy systems}

\author{K.~Lukierska-Walasek}
 \email{klukie@proton.if.uz.zgora.pl}
\affiliation{Institute of Physics, University of Zielona G\'ora, 
ul. Z. Szafrana 4a, 65-516 Zielona G\'ora, Poland}

\author{K.~Topolski}
\email{topolski@math.uni.wroc.pl}
\affiliation{Institute of Mathematics,
Wroc\l aw University, Pl.~Grunwaldzki 2/4, 50-384 Wroc\l aw, Poland
}

\begin{abstract}
We show that $p-$adic analysis provides a quite natural basis for the description of 
relaxation in hierarchical glassy systems. For our purposes, we specify the 
Markov stochastic process 
considered by S.~Albeverio and W.~Karwowski.
As a result we have obtained a random walk on $p-$adic
integer numbers, which provide the generalization of Cayley  tree proposed by Ogielski 
and Stein.  
The temperature-dependent power-law decay and the Kohlrausch law are derived.

\end{abstract}

\pacs{67.40.Fd, 75.10.Nr}
\keywords{Spin glass, dynamics of relaxation, $p-$adic, Markov processes}
\maketitle

\section{\label{sec:level1}Introduction}
The growing interest in  spin glassy dynamics has stimulated intense 
research in experimental \cite{1,2,3,4} as well as theoretical domain
\cite{5,6,7,8,9,10,11,12,13,14,15,16,17}. 
The theoretical studies have been successfully
developed in the following two directions:
\begin{description} 
\item[(i)] scaling theory for growing domains and droplets \cite{5,6,7}
\item[(ii)] the hierarchical structures in ultrametric spaces for relaxation dynamics
 \cite{8,9,10,11,12,13,14,15,16,17}.
\end{description}
The goal of this paper is to investigate the second item (ii), using $p-$adic analysis. 
The temperature cycle experiments \cite{1,2,3,4} are usually interpreted  
by the existence of a continuous hierarchy in spin glass dynamics. The hierarchical 
organization of metastable states depends on the temperature. It is assumed that
 at given temperature $T$
the system occupies restricted part of the phase space inside one of the free 
energy valleys. After cooling down to temperature $T-\Delta T,$  
where $\Delta T$ is rather small in the range 
$1^{\circ} K - 2^{\circ}K, $  such a part of phase space is divided into several smaller
descendent valleys separated by energy barriers. 
If the temperature is lowered the system has its new landscape of valleys but the
 information about the previous states is not lost {\em (memory effect).} 
So, after heating the system to 
temperature $T$ all descend states merge together into much smaller numbers 
of their ancestor states. 
The system in temperature $T$ continues  the
relaxation process.
After heating the system to critical temperature $T_c$ all occupied states 
merge into unique common ancestor state and phase space looks as in paramagnetic phase.

From temperature cycling experiments  it can also be seen, that the relaxation process at 
temperature $T $ during time $t$ is practically the same as the process at 
temperature $T' =T-\Delta T$ during some shorter time  $t'<t.$ 
The free energy barriers 
separating valleys at temperature $T'=T-\Delta T$ are higher than those at 
the temperature $T$ and corresponding amount of time $t$ needed to occupy the respective part 
of the phase space at temperature $T$ is smaller than  in the case of temperature 
$T'.$
Hierarchical structure of metastable states corresponds to such structure expressed in 
terms of pure states in the sense of Parisi solution \cite{17}.
If we lower the temperature the energy barriers  separating states are  becoming higher, 
even infinite. 

The dominant process in the relaxation dynamics has been described as the hopping between the 
physical states \cite{4}. Such mechanism leads to the studies of the models with a 
hierarchy-constrained dynamics 
with ultrametric topology.
These models did appear very useful for the description of complex systems, called 
 glassy 
systems, with highly degenerate metastable states.

\begin{figure}[h]
\includegraphics{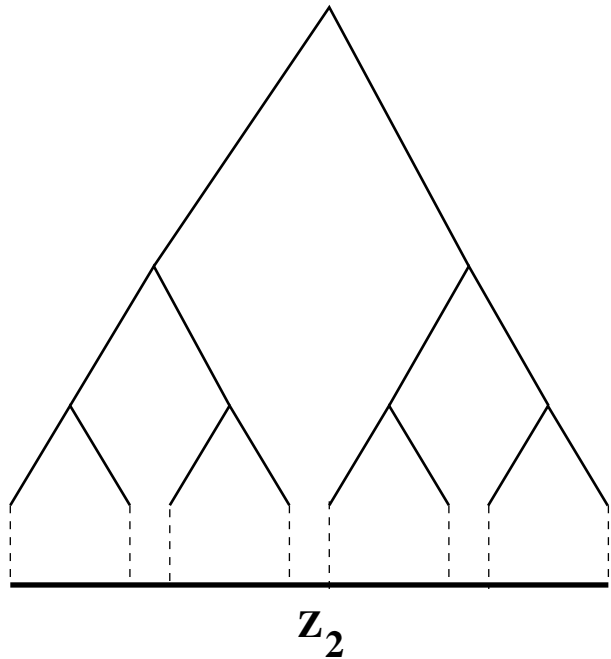}
\caption{\label{fig} Cayley tree with p=2. The bottom represents ${\bf Z_2}.$\\}
\end{figure}

In this paper we consider dynamics for glassy systems using a random walk on p-adic numbers. 
We specify, to our purposes, the 
Markov stochastic process, 
considered by S.~Albeverio and W.~Karwowski \cite{31,32}, for the description of dynamics 
in ultrametric
spaces. We use the space of  $p-$adic integers, which for our purposes 
is an appropriate 
example of 
ultrametric space. We should notice that  the $p-$adic integers may be  
represented by the bottom of the regular, infinite 
Cayley tree. Let us recall that in Cayley tree of order $p$ every branch at each 
level splits into $p$ other branches. For the illustration we present on Fig.~\ref{fig} 
the Cayley tree with $p=2$ from which we obtain, after infinity number of  splitting,
the {\em 2}-adic integers as describing the bottom of the tree. \\
We can introduce  a random walk in $p-$adic integers ${\bf Z_p}$ 
analogously as a random walk in a ultrametric 
space represented
by leaves of regular, finite Cayley tree.
The pure physical states of the system are represented by these leaves of  infinite 
Cayley tree. In order to construct the Markov process on 
{\em p}-adic integers we firstly consider Markov process 
on {\em p}-adic balls with the same finite radius and finally we contract the ball 
radius to zero.
In that way we obtain transition probabilities in the space of $p-$adic integers which 
correspond to the ones considered by Ogielski and Stein \cite{9}, in a case of finite 
ultrametric 
space.

The plan of the paper is the following. In Section 2 we introduce basic mathematical notions 
and the properties of 
$p-$adic integers. We describe as well a Markov process on {\em p-}adic 
integers.  In Section 3 we pass to the physical application.
We consider the thermally activated random walk on ultrametric space described as a Markov 
process on the space of {\em p-}adic balls. The temperature-dependent power-law decay 
and the Kohlrausch law are derived.
Let us notice finally that recently one observes an increasing interest in the application of 
p-adic numbers
in mathematical physics \cite{29}. The  p-adic 
analysis was used to  study 
stochastic processes \cite{Evans,29}, especially to ultrametric jump diffusion 
\cite{33,34,35}.
 
\section{Construction of Markov process on the $\bm p$-adic integers.}
Before we consider a Markov process with $p-$adic integers as a state space let us 
first introduce some notation and basic properties of $p-$adic numbers. More details 
may be fined, for example in \cite{koblitz} or \cite{29}.  
\subsection{$\bm p-$adic integers}
Let $p$ be an arbitrary  prime number and let ${\bf Z_p}$ denote the set 
of $p-$adic integers.
A $p-$adic integer is a {\em formal series} $\sum_{i\geq 0}a_i\,p^i$ 
{\em with coefficients} $a_i$
satisfying
$0\leq a_i \leq p-1.$ 
With this definition, a $p-$adic integer $a=\sum_{i\geq 0}a_i\,p^i$ 
can be identified with the 
sequence $(a_i)_{i \geq 0}$ of its coefficients.\\
In order to introduce a distance between $p-$adic integers $a$ and $b$ let us  first 
consider {\em an order} of a $p-$adic integer.
The {\em order} of a $p-$adic integers $a=(a_i)_{i \geq 0}$ is the smallest $m$
for which $a_m\not =0$
\begin{equation}
ord_p(...a_2\,a_1\,a_0)=\min\{i\,:\,a_i\not =0\},
\end{equation}
with the convention that maximum of the empty set is equal infinity.\\
Notice that if x, y are $p-$adic integers then 
\begin{subequations}
\label{allequations} 
\begin{eqnarray}
ord_p(xy)=ord_p(x)+ ord_p(y),\label{equationa}
\\
ord_p(x+y)\geq \min (ord_p(x),\, ord_p(y)).\label{equationb}
\end{eqnarray}
\end{subequations}
Now, in term of the function $ord_p(\cdot),$ we may introduce in the  space of $p-$adic 
integers the norm $|| \cdot ||_p$  
\begin{equation}
||x||_p=p^{-ord_p(x)},
\end{equation}
and  the $p-$adic metrics 
\begin{equation}
d_p(x,\,y)=||x-y||_p=p^{-ord_p(x-y)}.
\end{equation}
It is easy to check that the function $d_p$ is a metric, and moreover due to 
inequality (\ref{equationb}), fulfils
the following ultametric condition
\begin{equation}
d_p(x,y)\leq \max(d_p(x,z),\,d_p(z,y)).
\end{equation}
For $M\geq 0$ and $a\in {\bf Z_p}$ we may define a closed $p-$adic ball $B(a,M)$ 
with center $a$ and radius $p^{-M}$
\begin{equation}
B(a,M)=\left \{x\in {\bf Z_p} \,\,:\,d_p(x,a)\leq p^{-M}\right \}.
\end{equation}
If centre of the ball $a$ has $p-$adic representation
$a=\sum_{j=0}^{\infty}a_{j}p^{-j},$
then the ball $B(a,M)$ is completely determined by
\begin{equation}
\{a\}_{M}\equiv a_{M},\,a_{M-1},\dots,\,a_{0}.
\end{equation}
Let us recall some basic properties of the $p-$adic balls:
\begin{description} 
\item[(i)] for any $p-$adic balls  $B(a,M)$ and $B(b,M)$ we have
\begin{equation}
B(a,M)\cap B(b,M)=\emptyset\,\,\,\mbox{or}\,\,\,  B(a,M)=B(b,M),
\end{equation} 
which means that any $p-$adic balls are disjoint, or one is enclosed 
in another.
\item[(ii)]every point of the ball is the centre of this ball
\begin{equation}
\mbox{if}\,\,\,x\in B(a,\,M)\,\,\, \mbox{ then }\,\,\, B(x,\,M)=B(a,\,M).
\end{equation}
\item[(iii)] each ball $B(a,\,M)$ of  radius $p^{-M}$ may be represented as a 
finite union of {\bf disjoint}
balls $B(a_i,M+1)$ of radius $p^{-(M+1)}$
\begin{equation} 
B(a,\,M)=\bigcup_{i=0}^{p-1}B(a_i,\,M+1),
\end{equation}
\item[(iv)] (iii) implies that for {\bf disjoint}
balls $B(a_i,\,M)$ of radius $p^{-M}$
 \begin{equation}
{\bf Z_p}=\bigcup_{i=0}^{p^{M}-1}B(a_i,\,M).
\end{equation}
\end{description}
Let us finally define the distance between two arbitrary balls $B_1$ and $B_2$ as
\begin{equation}
d_p(B_1,B_2)= \inf \{d_p(x,\,y)\, : \,x\in B_1,\,\,y \in B_2\}.
\end{equation}
With such definition two balls with representations 
$\{a\}_{M}\equiv a_{M},a_{M-1},\dots ,a_k,a_{k-1},\dots,a_{0}$
and $\{b\}_{M}\equiv b_{M},b_{M-1},\dots ,b_k,a_{k-1},\dots,\,a_{0}$, 
differ for the 
first time at 
$k-th$ position,  we have
\begin{equation}
d_p(\{a\}_{M},\,\{b\}_{M})= p^{-k}.
\end{equation}
 
\subsection{Random walk on $\bm{p-}$adic balls}
Let $\{X(t),\,t\geq 0\}$ be a Markov process with the state space ${\bf Z_p}$ and 
transition rates between states depending only on $p-$adic distance between these states.
To define such a process we a follow  general construction given in \cite{32}.
In order to do this we represent ${\bf Z_p}$   
as a finite union of disconnected 
balls $B_i^{M}$ with the radius $p^{-M}$ 
\begin{equation}
{\bf Z_p}=\bigcup_{i=0}^{p^{M}-1}B_i^{M}
\end{equation}
For a fixed $M$ we may consider a finite state space Markov process $\{X_M(t),\,t\geq 0\}$  
with the state space 
$E_M=\{B^{M}_0,\, B^{M}_1,\,\dots,\,B^{M}_{p^{M}-1}\},$ set of $p-$adic balls with the
 radius $p^{-M},$    
and transition probabilities  from ball $B_i^{M}$ at time $0$ to ball  
$B_j^{M}$ at time $t,$ $P^{(M)}_{i,j}(t),$ defined as
\begin{eqnarray}
P^{(M)}_{i,j}(t)&\equiv& P\left(X_M(t)= B_j^{M}\,|\,X_M(0)= B^{M}_i\right)\nonumber\\
\nonumber\\
&=&P\left(X(t)\in B_j^{M}\,|\,X(0)\in B^{M}_i\right).
\end{eqnarray}
Transition probabilities 
$P^{(M)}_{i,j}(t),\,(i,j=0,1,\dots,p^{M}-1)$ are 
solutions of the
system of Kolmogorov equations
\begin{equation}
\frac{d}{dt}\,P^{(M)}_{i,j}(t)=-q^{(M)}_i\,P^{(M)}_{i,j}(t)+\sum_{0\leq k \leq p^M-1
 \atop k \not=i}
q^{(M)}_{ik}\,
P^{(M)}_{k,j}(t),
\end{equation}
with the initial condition $P_{i,j}(0)=\delta _{ij},$
where $q^{(M)}_{ij}$ the infinitesimal transition probability and $q^{(M)}_i$ the
 intensity of stay in state $i$ of the process are defined 
for any pair $i,\,j$ of different states represented by balls $B^{M}_i,\,B^{M}_j$
 with radius $p^{-M}$, by
\begin{equation}
q^{(M)}_{ij}=\lim_{h\downarrow 0} \frac{P^{(M)}_{i,j}(h)}{h},
\end{equation}
and for any state $i,$ by
\begin{equation}
q^{(M)}_{i}=\lim_{h\downarrow 0} \frac{1-P^{(M)}_{i,j}(h)}{h}.
\end{equation}

Observe, that each ball $B^{M}_i$ is a union of disjoint balls $B^{M+1}_{ik},$
$(k=1,\,\dots,p)$ of 
radius $p^{-(M+1)}$. We stress here, that $q^{(M)}_{ij}$ depend only on 
$p-$adic distance between balls $B^{M}_i$
and $B^{M}_j.$ 
For this reason we may represent transition probabilities of the process $X_M(t)$
 by appropriate
transition probabilities of the process $X_{M+1}(t)$
\begin{eqnarray}
P^{(M)}_{i,j}(t)&=&P\left(X_M(t)= B_j^{M}\,|\,X_M(0)= B_i^{M}\right)\nonumber\\
\nonumber\\
&=&P\left(X(t)\in \bigcup_{k=1}^{p}B_{jk}^{M+1}\,|\,X(0)\in 
\bigcup_{k=1}^{p} B_{ik}^{M+1}\right)\nonumber\\
\nonumber\\
&=&pP\left(X_{M+1}(t)= B_j^{M+1}\,|\,X_{M+1}(0)= B_i^{M+1}\right)\nonumber\\
\nonumber\\
&=&pP^{(M+1)}_{i,j}(t))\label{req}
\end{eqnarray}
This equality leads to the recurrence relation for the local 
characteristics 
$q^{(M)}_{i,j}$
and $q_{i,j}^{(M+1)}$
 of the 
processes $\{X_M(t)\}$ and $\{X_{M+1}(t)\}.$
Let $dist_p(B^M_i,\,B^M_j)=p^{-n},$ then we define 
\begin{equation}
\label{int}
q^{(M)}_{i,j}\equiv p^{-M}u(-M,M-n).
\end{equation}
we may notice that definition of $q^{(M)}_{i,j}$ and (\ref{req}) impies
\begin{equation}
\label{equ1}
u(-M+1,m-1)=u(-M,m).
\end{equation}
Finally taking into account that $p^{-M}u(M,m)$ represents probability intensity  transition 
of the
Markov process 
and (\ref{equ1}) we may write
\begin{equation}
\label{a}
u(-M,m)=a(-M+m-1)-a(-M+m),
\end{equation}
where $\{a(-n)\},$ $n=0,\,1,\,2,\,\dots.$ is a sequence of positive numbers such that
\begin{equation}
a(-n)\geq a(-n+1)
\end{equation}

Proceeding in the similar way as in \cite{32} 
we obtain the solution of the Kolmogorov equations  for the Markov process with local
 characteristics 
 $q_{i,j}^{(M)}$ given by (\ref{int}) and (\ref{equ1}) in the form
\begin{eqnarray}
\label{trans1d}
P^{(M)}_{i,i}(t)=p^{-M}+\frac{p-1}{p}\sum_{i=0}^{M-1}p^{-i}
\exp(t{\cal{W}}_{-M,i+1}),
\end{eqnarray}
if $dist_p(B^{M}_i,\,B^{M}_j)=p^{-M+m}$ then 
\begin{eqnarray}
P^{(M)}_{i,j}(t)=p^{-M}+\frac{p-1}{p}
\sum_{i=0}^{M-1}p^{-(m+i)}\exp(t{\cal{W}}_{-M,m+i+1})
-p^{-m}\exp(t{\cal{W}}_{-M,m})
\end{eqnarray}
where 
\begin{equation}\label{W}
{\cal{W}}_{-M,j}=-\sum_{k=j}^{M-1}(u(-M,k)-u(-M,k+1))p^{-M+k}.
\end{equation}
Now we define the 
transition probabilities of the 
Markov process on ${\bf Z_p}$ in the following way.\\
For any $x \in {\bf Z_p} $ let 
\begin{eqnarray}
\label{trans2}
P_t(x;\,B(a,M))\equiv
P\left(X_M(t)\in B(x,M)\,|\,X_M(0)\in B(a,M)\right),
\end{eqnarray}
By arguments  similar to those in \cite{32} we
may prove that there exists continuous time
Markov stochastic process $\{X(t),\,t\geq 0\}$ 
with state space ${\bf Z_p}$ 
and transition probabilities $P_t(x;\,B)$ given by  
(\ref{trans2}).\\
Observe that (\ref{trans2}) give us direct connection between Markov processes 
$\{X_{M}(t),\,t\geq 0\}$ on $p-$adic balls with radius $p^{-M},$ and Markov
process $\{X(t),\,t\geq 0\}$ on ${\bf Z_p}.$   
\section{Dynamics as a thermal hopping in $p-$adic space.}
In this section we shall apply the mathematical considerations from Section 2 to the physical 
system with the hierarchy of states which can be linked with Cayley tree structure. 
By studying the dynamics of such the systems temperature-dependent power law decay 
and the Kohlrausch law are derived.
Transitions between states are thermally activated. The height of the energy barriers,
 $\Delta_k\,$ $k=1,\,2,\,\dots,$ which the system  overcomes can be ordered in the 
increasing sequence 
$\Delta_1 < \Delta_2 < \dots < \Delta_k \dots.$   The time evolution of the system  
is described 
by a random walk on the space of states.
The simplest model of  such dynamics is the model proposed by Ogielski and Stein \cite{9}.
They consider a regular  Cayley tree with $M$  levels and fixed branching ratio $p.$ 
The total number of 
 leaves, points on the bottom of  the tree,  is $n=p^M.$  The natural ultrametric 
distance $d(k,\,l)$ 
between leaves $k$ and $l$ , is defined as equal to the height 
$m,\,\,m=0,\,1,\,\dots,\,M$ of their closest common ancestor. Now, identifying the 
states $x$ and $y$ 
separated by the energy barrier $\Delta_m$  with leaves $k$ and $l,$ the probability of 
moving from state $x$
 to state $y$ may be defined as equal to transition probability from  leaf  $k$ to
 leaf  $l$ separated by the ultrametric distance
 $m.$ Thus dynamics in the space of states
separated by the energy barrier may be studied in terms of appropriate Markov process 
involving the  end points of 
Cayley tree, as a space of states. It is a nontrivial observation that the probability 
transition intensities between states are
depending on their ultrametric 
distance. Due to hierarchical structure of the state space a probability intensities  
matrix of the process has 
Parisi matrix structure. Parisi matrix has regular form, and  for illustration we 
present the case $p=2.$
\begin{equation}
\left [ 
\begin{array}{cccc}
\begin{array}{cc}
\begin{array}{cc}
\epsilon_{0}&\epsilon_{1}\\
\epsilon_{1}&\epsilon_{0}\\
\end{array}& {\bf E_1}\\
{\bf E_1}&
\begin{array}{ccc}
\epsilon_{0}&\epsilon_{1}\\
\epsilon_{1}&\epsilon_{0}\\
\end{array}
\end{array}&{\bf E_2}& &\\
{\bf E_2}&\begin{array}{cccc}
\begin{array}{cc}
\epsilon_{0}&\epsilon_{1}\\
\epsilon_{1}&\epsilon_{0}\\
\end{array}& {\bf E_1}&\\
{\bf E_1}& 
\begin{array}{cccc}
\epsilon_{0}&\epsilon_{1}& &\\
\epsilon_{1}&\epsilon_{0}& &\\
\end{array}
\end{array}\\
\multicolumn{4}{c}{\dotfill}
\end{array}\\ 
\right ],
\end{equation}
where $\bm E_i$ is the matrix 
with all elements equal to $\epsilon_i.$\\
One can  observe that end points of  a regular  Cayley  tree with $M$  levels 
and fixed branching ratio $p$  may be represented as a set  of disconnected 
balls $\{B_0^{M},\,\dots,\,B_{p^{M}-1}^{M}\}$ with  radius 
$p^{-M}$ covering $\bm Z_p.$

Let us consider now a  special case of Markov process on $p-$adic integers. 
We assume that transition probability intensities of the process $\{X(t),\,t\geq 0\}$  
depend on $p-$adic distance only. For this process we have a corresponding 
Markov chain $\{X_M(t),\,t\geq 0\}$
with the set of disconnected 
balls $\{B_0^{M},\,\dots,\,B_{p^{M}-1}^{M}\}$ with the radius $p^{-M}$ covering 
$\bm Z_p,$ as a state space. If we enumerate these balls in such a way that 
$dist_p(B_0^{M},\,B_{i}^{M})$
increase with $i$ then  ${\bf Q}=[q_{ij}],$ $(0\leq j \leq p^M-1,$ $0\leq i \leq p^M-1),$
 the matrix of 
transition probability intensities of the process $\{X_M(t),\,t\geq 0\}$ 
has Parisi matrix form.  By appropriate choice of $u(-M,\,k),$ 
we obtain process studied in \cite{9}.\\
Let for $k=1,\,2,\,\dots,\,M-1$
\begin{equation}\label{epsilon}
\epsilon_k=p^{-M}u(-M,\,k)\equiv q_{ij}^{(M)},
\end{equation}
where $q_{ij}^{(M)}$ is probability transition intensity of a jump from the ball 
$B_{i}^{M}$ to the ball $B_{j}^{M},$ separated by $p-$adic distance $p^{-(M-k)}.$\\ 
\begin{widetext}
From (\ref{trans1d}) we have that process which at time $0$ starts from a ball $B_0^M$ 
will be found at this ball at time $t$  with probability 
\begin{eqnarray}
\label{termal}
P_t(B_0^M,\, B_0^M)=p^{-M}+\frac{p-1}{p}\sum_{i=1}^{M-1}p^{-i}
\exp(-t{\cal{W}}_{-M,i+1})\}
\end{eqnarray}
which together with (\ref{W}) and (\ref{epsilon}) gives
\begin{eqnarray}
\label{alg}
P_t(B_0^M,\,B_0^M)=
p^{-M}+\frac{p-1}{p}\sum_{i=0}^{M-1}p^{-i}
\exp(-t\sum_{k=i+1}^{M}(\epsilon_{k}-\epsilon_{k+1})p^{k})
\end{eqnarray}
For a special case $p=2$ equation (\ref{alg}) has the same form as corresponding
 equation (6) from
\cite{9}
\begin{equation}
P_t(B_0^M,\, B_0^M)=2^{-M}+\frac{1}{2}\sum_{i=0}^{M-1}2^{-i}
(\exp(-t[2a_{i+1}+\sum_{k=i+2}^{M}a_k])
\end{equation}
where $a_k=2^{k-1}\,\epsilon_k$ represents the probability intensity of a 
jump an ultrametric distance $k$ from a starting sit at Cayley tree, while in our
 case $a_k$ represents the probability transition intensity of a
jump to 
any ball of radius
$2^{-M}$ at $2-$adic distance $2^{-(M-k)}.$ 
\end{widetext}
Finally we contract the ball radiuses to zero and performing procedure analogical
 to \cite{9} and 
specifying the form of energy barriers and probability intensities of crossing a barriers
we are able to compare different scenario.  
The most simple case is a sequence of barriers linearly growing with $p-$adic 
distance $\Delta_k=\Delta\, k$ for some positive constant $\Delta$   and
$a_k=e^{-\Delta\,k/T}$ and for fixed temperature $T.$ 
In this case (\ref{alg}) gives us fore large time t a temperature-dependent power law.
\begin{equation}
\lim_{M\to\infty} P_t(B_0^M, B_0^M) \sim t^{-T\,ln\,2/\Delta}.
\end{equation}
For a sequence of energy barriers which grows in slower way, i.e $\Delta_k=\Delta\,\ln k,$
 for some 
positive constant $\Delta$ 
and 
$a_k=e^{-\Delta\,\ln k/T}$ 
and for fixed temperature $T,$  the probability $P_t(B_0^M,\,B_0^M)$
fulfils,  for large $t,$ the Kohlrausch law
\begin{equation}
\lim_{M \to \infty}P_t(B_0^M, B_0^M) \sim \exp(-t^{T/\Delta}).
\end{equation}

\section{Concluding Remarks}
In our paper use random walk framework and we show that $p-$adic analysis is very 
natural tool to describe the relaxation process in a glassy systems. $p-$adic space 
has a well defined ultrametric topology which we employ in this paper. 
We do not use the random walk along the Cayley  tree \cite{9} because  
one can measure the ultrametric distance between the physical states 
directly on the bottom of the tree.

For the description of dynamics of the hierarchical systems besides of 
the random walk on $p-$adic numbers one can alternatively use the process 
of jump diffusion on $p-$adic numbers which description employs the pseudodifferential 
operators \cite{koch} in $p-$adic space. In these both approaches one obtains the 
equivalent physical results \cite{33,34,35}.

It seems that the $p-$adic analysis is a good tool not only for the description 
of the dynamics of the glassy systems, but also for another hierarchical processes 
like the evolution of  fractals \cite{frac}, the avalanches \cite{aval}, 
protein folding \cite{prot} ect.. It is interesting to notice that the 
$p-$adic space inherently includes the natural  hierarchy: $p-$adic balls 
can be covered by the smaller ones disjoint balls. The hierarchy of these 
nested balls correspond to the hierarchy of the scales of the configuration 
rearrangements, as it is seen from the scaling theory for the growing domains 
and droplets [5--7]. The droplets may be broken into smaller ones when 
temperature decrease.

It is also interesting to notice that by the mapping of any Cayley tree 
onto its bottom states one can obtain the integer $p-$adic space and 
further changing the fractional part of the main ancestors  we obtain 
another integer $p-$adic space. $p-$adic space consists of infinite 
number of these integer $p-$adic spaces. In conclusion one can comment 
that the  memory effects in spin glassy systems it seems to be are well
described by  $p-$adic topology.  
\begin{acknowledgments}
We are grateful to Prof. Witold Karwowski for his interest in this paper
and valuable comments.
\end{acknowledgments}

\bibliographystyle{apsrev}
  
\end{document}